\documentclass[12pt]{iopart}

\usepackage{newtxtext}
\usepackage{comment}
\usepackage{cite}
\usepackage{amssymb}
\usepackage{amsthm}
\usepackage{iopams}

\theoremstyle{definition}
\newtheorem{Def}{Definition}[section]

\newtheorem{rem}[Def]{Remark}

\newcommand{\const}{\mathrm{const}.}
\newcommand{\al}{\alpha}
\newcommand{\be}{\beta}
\newcommand{\del}{\delta}
\newcommand{\eps}{\epsilon}

\newcommand{\bfrac}[2]{\left(\frac{#1}{#2}\right)}
\newcommand{\od}[3][1]{
\ifnum #1=1\frac{\mathrm{d} #2}{\mathrm{d} #3}
\else\frac{\mathrm{d}^{#1} #2}{\mathrm{d} #3^{#1}}
\fi}
\newcommand{\pd}[3][1]{
\ifnum #1=1\frac{\partial #2}{\partial #3}
\else\frac{\partial^{#1} #2}{\partial #3^{#1}}
\fi}
\newcommand{\hd}[4]{D^{#1}_{#2}{#3}\cdot{#4}}

\newcommand{\dlydiff}{delay-differential}

\newcommand{\pain}{Painlev\'e}
\newcommand{\lv}{Lotka-Volterra}
\newcommand{\toda}{Toda lattice}
\newcommand{\sg}{sine-Gordon}

\newcommand{\kp}{KP equation}

\begin{document}

\title[A systematic construction of integrable
delay-analogues of soliton equations]
{A systematic construction of integrable
delay-difference and delay-differential analogues of soliton equations}
\author{Kenta Nakata$^1$ and Ken-ichi Maruno$^2$
}
\address{$^1$~Department of Pure and Applied Mathematics, School of Fundamental Science and Engineering, Waseda University, 3-4-1 Okubo, Shinjuku-ku, Tokyo 169-8555, Japan
}
\address{$^2$~Department of Applied Mathematics, Faculty of Science and Engineering, Waseda University, 3-4-1 Okubo, Shinjuku-ku, Tokyo 169-8555, Japan
}
\ead{kennakaxx@akane.waseda.jp and kmaruno@waseda.jp}
\date{\today}

\vspace{10pt}
\begin{indented}
\item[]{\today}
\end{indented}

\begin{abstract}
We propose a systematic method for constructing integrable delay-difference and \dlydiff\ analogues of known soliton equations such as the \lv, \toda, and sine-Gordon equations and their multi-soliton solutions.
It is carried out by applying a reduction and delay-differential limit to the discrete KP or discrete two-dimensional Toda lattice equations.
Each of the delay-difference and \dlydiff\ equations has the $N$-soliton solution, which depends on the delay parameter and converges to an $N$-soliton solution of a known soliton equation as the delay parameter approaches $0$.
\end{abstract}

\submitto{\jpa}

\begin{section}{Introduction}

Delay-differential equations have been used as mathematical models
in various fields of science and engineering such as traffic flow, population dynamics, nonlinear optics, fluid mechanics and infectious disease~\cite{delaybook,delaybiologybook}.
Because of their importance in various fields,
the mathematical properties of delay differential equations have been actively studied.
For example, exact solutions of delay differential equations play an important role in
the study of traffic flow~\cite{nakanishi,hasebe,kanai,matsuya}.

In recent years, there has been active research on integrability of
delay differential equations.
Quispel \textit{et al} obtained a \dlydiff\ equation which has a continuum limit to the first \pain\ equation~\cite{Quispel}.
This \dlydiff\ equation was derived by a similarity reduction of the Lotka-Volterra (LV) equation, which is an integrable differential-difference equation.
Levi and Winternitz also obtained delay-differential equations
as reductions based on the symmetry group of the two-dimensional Toda lattice (2DTL)~\cite{Levi}.
In addition, Grammaticos \textit{et al}~\cite{Gram,Ramani} introduced delay-differential \pain\ equations by the method blending the \pain\ analysis with the singularity confinement.
After their monumental works, several important studies
have revealed mathematical properties
of integrable (ordinary) \dlydiff\ equations~\cite{Joshi1,Joshi2,Carstea,Viallet,Halburd,Berntson,Stokes}.

On the other hand, there are few examples of
integrable partial \dlydiff\ equations (the word ``partial'' means including multi independent variables).
Villarroel and Ablowitz found the Lax pair
for a \dlydiff\ analogue of the 2DTL equation and
established the inverse scattering formalism~\cite{ablowitz}.
Recently, we constructed the $N$-soliton solution of the \dlydiff\ analogues
of the 2DTL equation~\cite{delay2DTL}.
Although there have been such works, research on integrable partial \dlydiff\ equations is not fully developed.

In this paper, we propose a systematic method for constructing
integrable delay-difference and
delay-differential analogues of soliton equations and their multi-soliton solutions.
Our construction starts from the discrete KP equation (or discrete 2DTL equation)
and uses reduction and delay-differential limit.
As examples of our method, we present delay-difference and
delay-differential analogues of the LV, \toda\ (TL), and
\sg\ (sG) equations and their exact $N$-soliton solutions.

This paper is organized as follows.
In the rest of this section, we show a systematic method to construct integrable delay-difference and delay-differential analogues of soliton equations.
In sections 2, 3, and 4, we construct delay-difference and delay-differential analogues of the LV, TL, and sG equations and their $N$-soliton solutions.
Section 5 is devoted to conclusions.

\begin{subsection}{A method to construct delay-analogues of soliton equations}

In the rest of this section, we show how to construct integrable delay-difference and delay-differential analogues of soliton equations and their $N$-soliton solutions.
The first step of our method is to obtain a discrete equation by a reduction of the discrete KP equation~\cite{Hirota-discrete_KP,Miwa-discrete_KP}
\begin{eqnarray}
    \label{dkp}
     a(b-c)f_{n+1,m,k}f_{n,m+1,k+1}
    +b(c-a)f_{n,m+1,k}f_{n+1,m,k+1}\nonumber\\
    \hspace{20mm} +c(a-b)f_{n,m,k+1}f_{n+1,m+1,k}
    =0
\end{eqnarray}
or the discrete 2DTL equation~\cite{Hirota-discrete_KP,Hirota-discrete2DTL}
\begin{eqnarray}
    abf_{k+1,n+1,m}f_{k-1,n,m+1}
    +f_{k,n+1,m}f_{k,n,m+1}\nonumber\\
    \hspace{20mm} -(1+ab)f_{k,n+1,m+1}f_{k,n,m}=0\,,\label{2dtl}
\end{eqnarray}
where $a, b, c$ are real constants.
Here we include a free parameter $\al$ to the reduction condition such as (\ref{reduction:LV}).
By applying the reduction, we obtain a discrete equation which depends on the free parameter $\al$, the discrete variable $n, m$, and the time-lattice parameter $\del$ which is defined by the parameters $a, b, c$ such as (\ref{dlydislv}).
Note that the dependency on $k$ is removed by the reduction, and the parameter $\al$ appears as shifts of $m$ (such as $f_{m+\al}$).
This discrete equation can be considered as a delay-difference soliton equation.

Then, we apply the delay-differential limit
\begin{equation}
    \del\to0\,,\qquad
    m\del=t\,,\qquad
    \al\del=\tau=\const
\end{equation}
to the delay-difference soliton equation, where $t$ is the continuous time variable and $\tau$ is the delay parameter.
This limit yields a \dlydiff\ equation which includes the delay parameter $\tau$ as shifts of $t$ such as (\ref{dlylv}).
We can actually check that the limit of $f_{m+\al}$ is $f(t+\tau)$.
This \dlydiff\ equation can be considered as a \dlydiff\ soliton equation.

An important point of this process is that we can obtain explicit $N$-soliton solutions of the delay-difference and \dlydiff\ soliton equations.
It is carried out by reduction and delay-differential limit to the $N$-soliton solution of the discrete KP or discrete 2DTL equations.
In addition, we can obtain a known soliton equation and its $N$-soliton solution as $\tau\to0$ if we properly determine the reduction at the first step of our method.
Here the soliton equation which is obtained as $\tau\to0$ depends on the reduction.
Therefore, considering various reductions at the first step of our method, we can construct various delay-difference and delay-differential analogues of soliton equations.

\end{subsection}

\end{section}

\begin{section}{An integrable delay \lv\ equation}

In this section, we show the construction of the delay-difference and delay-differential analogue of the LV equation and their $N$-soliton solutions.
We show the detail of our method through this section.

First, we consider the discrete KP equation (\ref{dkp}) and
its $N$-soliton solution in the Gram determinant form~\cite{Ohta-discrete_KP}:
\begin{eqnarray}
\label{dkp_sol}
&&\fl f_{n,m,k}=\det\left(
\delta_{ij}+\frac{\phi_i\psi_j}{p_i-q_j}
\right)_{1\leq i,j\leq N}
=\frac{\prod_{i=1}^N \phi_i}{\prod_{j=1}^{N} \phi_j}
\det \left(\delta_{ij}\frac{\phi_j}{\phi_i}+\frac{\phi_j \psi_j}
{p_i-q_j}\right)_{1\leq i,j\leq N}\\
&&\fl \hspace{1cm} =
\det \left(\delta_{ij}+\frac{\phi_j \psi_j}
{p_i-q_j}\right)_{1\leq i,j\leq N}\,,\nonumber\\
&&\fl \phi_i(n,m,k)=\beta_i(1-ap_i)^{-n}(1-bp_i)^{-m}(1-cp_i)^{-k}\,,\nonumber\\
&& \fl\psi_i(n,m,k)=\gamma_i(1-aq_i)^{n}(1-bq_i)^{m}(1-cq_i)^{k} \,.\nonumber
\end{eqnarray}
This Gram determinant form of the $N$-soliton solution can be rewritten as
\begin{eqnarray}
\label{dkp_sol2}
&&f_{n,m,k}
    =\sum_{I\subset\{1,\ldots,N\}}
    \prod_{i\in I}\Phi_i
    \prod_{i<j,\ i,j\in I}\frac{(p_i-p_j)(q_i-q_j)}{(p_i-q_j)(q_i-p_j)}\,,\\
&&\Phi_i(n,m,k)
    =\beta_i \gamma_i
\bfrac{1-aq_i}{1-ap_i}^{n}\bfrac{1-bq_i}{1-bp_i}^{m}\bfrac{1-cq_i}{1-cp_i}^{k}\,.\nonumber
\end{eqnarray}
Here $\displaystyle \sum_{I\subset\{1,\ldots,N\}}$ is taken to be summed over all possible
combinations in the set $\{1,\ldots,N\}$.
Now, we apply the following reduction condition to the discrete KP equation (\ref{dkp}):
\begin{equation}
    f_{n,m,k+1}=f_{n+2,m+\al,k}\,,\label{reduction:LV}
\end{equation}
where the free parameter $\al$ is a fixed real value considered as the delay.
Setting $c=-a$ and $\del=2b/(a-b)$, we obtain the following discrete bilinear equation by the reduction of the discrete KP equation (\ref{dkp}):
\begin{equation}
    \label{dlydislv}
    (1+\del)f_{n}^{m+1+\al}f_{n-1}^{m}
    -\del f_{n+1}^{m+\al}f_{n-2}^{m+1}
    -f_{n}^{m+\al}f_{n-1}^{m+1}
    =0\,,
\end{equation}
where $f_n^m\equiv f_{n,m,k}$.
More precisely, applying the reduction condition (\ref{reduction:LV}) to equation (\ref{dkp}), we have made the index $k+1$ of $f$ become $k$.
Then we can omit the variable $k$ since the iterations of $k$ vanish.
We can rewrite the bilinear equation (\ref{dlydislv}) as follows by using Hirota's D-operators:
\begin{eqnarray}
    \fl\left(2\sinh\bfrac{D_m}{2}\sinh\left(\frac{D_n}{2}+\frac{\al D_m}{2}\right)
    -2\del\sinh\bfrac{D_n-D_m}{2}\sinh\left(D_n+\frac{\al D_m}{2}\right)\right)
    f_{n}^{m}\cdot f_{n}^{m}\nonumber\\
    =0\,.
\end{eqnarray}
Here Hirota's D-operators are defined by
\begin{equation}
    \fl\hd{l}{t}{g(t)}{h(t)}
    =\left(\pd{}{t}-\pd{}{s}\right)^{l}g(t)h(s)\Bigr \vert_{s=t}\,,\qquad
    e^{D_m}g_{m}\cdot h_{m}
    =g_{m+1}h_{m-1}\,.
\end{equation}

Next we consider the reduction of the $N$-soliton solution.
By applying the constraint
\begin{equation}
    \bfrac{1-aq_i}{1-ap_i}^{2}\bfrac{1-bq_i}{1-bp_i}^{\alpha}=\frac{1-cq_i}{1-cp_i}
\end{equation}
to the $N$-soliton solution (\ref{dkp_sol}) and (\ref{dkp_sol2}),
we obtain
\begin{eqnarray}
\Phi_i(n,m,k+1)&=&\beta_i \gamma_i
\bfrac{1-aq_i}{1-ap_i}^{n}\bfrac{1-bq_i}{1-bp_i}^{m}\bfrac{1-cq_i}{1-cp_i}^{k+1}\nonumber\\
&=&\beta_i \gamma_i
\bfrac{1-aq_i}{1-ap_i}^{n+2}\bfrac{1-bq_i}{1-bp_i}^{m+\alpha}\bfrac{1-cq_i}{1-cp_i}^{k}
\nonumber\\
&=&\Phi_i(n+2,m+\alpha,k)\,,
\end{eqnarray}
thus the reduction condition (\ref{reduction:LV}) is satisfied.
Setting $c=-a$, $\del=2b/(a-b)$, $k=0$,
and replacing $p_i$ and $q_i$ by $(-2p_i-1)/a$ and $(-2q_i-1)/a$
respectively, we obtain the $N$-soliton solution of the bilinear equation (\ref{dlydislv})
by the reduction to (\ref{dkp_sol}) and (\ref{dkp_sol2}):
\begin{eqnarray}
\label{dlydislv_sol}
\fl && f_{n}^{m}
=
\det \left(\delta_{ij}+\frac{\Phi_j}
{p_i-q_j}\right)_{1\leq i,j\leq N}
=\sum_{I\subset\{1,\ldots,N\}}
    \prod_{i\in I}\Phi_i
    \prod_{i<j,\ i,j\in I}\frac{(p_i-p_j)(q_i-q_j)}{(p_i-q_j)(q_i-p_j)}\,,\\
\fl && \Phi_i(n,m)
    =\beta_i\gamma_i\bfrac{1+q_i}{1+p_i}^{n}
\bfrac{1+\del+\del q_i}{1+\del+\del p_i}^{m}\,,\nonumber\\
\fl && \frac{q_i}{p_i}
    =\bfrac{1+q_i}{1+p_i}^2\bfrac{1+\del+\del q_i}{1+\del+\del p_i}^{\al}\,.\nonumber
\end{eqnarray}
The delay parameter $\al$ appears in the bilinear equation (\ref{dlydislv})
as shifts of the discrete time variable $m$.
We remark that the bilinear equation (\ref{dlydislv}) and the $N$-soliton solution
(\ref{dlydislv_sol}) in the case of
$\al=0$ (thus $p_i=1/q_i$) are known as the bilinear equation of the
discrete LV equation and its $N$-soliton
solution~\cite{Hirota-discrete_soliton_equations,Maruno-discrete_soliton_equations}.
Thus equation (\ref{dlydislv}) can be seen as
the bilinear equation of the delay-difference analogue of the LV equation.

By the dependent variable transformation
\begin{equation}
    u_{n}^{m}
    =\frac{f_{n+1}^{m+\al}f_{n-2}^{m+1}}{f_{n}^{m+\al}f_{n-1}^{m+1}}\,,
\end{equation}
the bilinear equation (\ref{dlydislv}) is transformed into the nonlinear delay-difference
equation
\begin{equation}
    \label{dlydislv_nl}
    \frac{u_{n}^{m+1+\al}u_{n-1}^{m}}{u_{n}^{m+\al}u_{n-1}^{m+1}}
    =\frac{(1+\del u_{n+1}^{m+\al})(1+\del u_{n-2}^{m+1})}{(1+\del u_{n}^{m+\al})(1+\del u_{n-1}^{m+1})}\,,
\end{equation}
which is the delay-difference analogue of the LV equation.
If $\al=0$, equation (\ref{dlydislv_nl}) is just a division of
the discrete LV equation~\cite{Hirota-discrete_soliton_equations}:
\begin{equation}
    \frac{u_{n}^{m+1}}{u_{n}^{m}}
    =\frac{1+\del u_{n+1}^{m}}{1+\del u_{n-1}^{m+1}}\,,\qquad
    \frac{u_{n-1}^{m+1}}{u_{n-1}^{m}}
    =\frac{1+\del u_{n}^{m}}{1+\del u_{n-2}^{m+1}}\,.
\end{equation}

Now, we apply the \dlydiff\ limit
\begin{equation}
    \del\to0\,,\qquad
    m\del=t\,,\qquad
    \al\del=2\tau\,,
\end{equation}
where $\tau$ is a constant value called the delay parameter.
The \dlydiff\ limits of (\ref{dlydislv}) and (\ref{dlydislv_sol})
are calculated respectively as follows:
\begin{eqnarray}
    \label{dlylv}
    \fl\hd{}{t}{f_{n}(t+\tau)}{f_{n-1}(t-\tau)}
    -f_{n+1}(t+\tau)f_{n-2}(t-\tau)
    +f_{n}(t+\tau)f_{n-1}(t-\tau)
    =0\,,
\end{eqnarray}
\begin{eqnarray}
\fl && f_{n}(t)
=
\det \left(\delta_{ij}+\frac{\Phi_j}
{p_i-q_j}\right)_{1\leq i,j\leq N}
=\sum_{I\subset\{1,\ldots,N\}}
    \prod_{i\in I}\Phi_i
    \prod_{i<j,\ i,j\in I}\frac{(p_i-p_j)(q_i-q_j)}{(p_i-q_j)(q_i-p_j)}\,,\label{dlylv_sol}\\
\fl &&\Phi_i(n,t)
    =\beta_i\gamma_i\bfrac{1+q_i}{1+p_i}^{n}e^{(q_i-p_i)t}\,,\nonumber\\
\fl &&\frac{q_i}{p_i}
    =\bfrac{1+q_i}{1+p_i}^2e^{2\tau(q_i-p_i)}\,.\nonumber
\end{eqnarray}
Here the bilinear equation (\ref{dlylv}) can be rewritten as
\begin{eqnarray}
    \label{dlylv2}
    \fl\left(D_t\sinh\left(\frac{D_n}{2}+\tau D_t\right)
    -2\sinh\bfrac{D_n}{2}\sinh\left(D_n+\tau D_t\right)\right)
    f_{n}(t)\cdot f_{n}(t)=0\,.
\end{eqnarray}
The bilinear equation (\ref{dlylv}) is a delay differential-difference
equation which includes the delay $\tau$ as shifts of the continuous time variable $t$.
Putting $\tau=0$ (thus $p_i=1/q_i$), we can easily check that equation
(\ref{dlylv})
becomes the bilinear equation of
the LV equation and (\ref{dlylv_sol}) becomes the $N$-soliton solution
of the LV equation~\cite{Hirota-LV}.
Thus we claim that equation (\ref{dlylv})
is the bilinear equation of the integrable delay
LV equation and the bilinear equation (\ref{dlydislv}) is
the fully discrete analogue of (\ref{dlylv}).

\begin{rem}
The last relation of (\ref{dlylv_sol}), which is the dispersion relation,
is rewritten as
\begin{equation}
    R(p_i)=R(q_i)\,,\qquad R(p)=\frac{pe^{-2\tau p}}{(1+p)^2}\,.
\end{equation}
According to the graph of $R(p)$, it is easily seen that
there exist $p_i$ and $q_i$ satisfying $R(p_i)=R(q_i)$ and $p_i\neq q_i$.
\end{rem}

\begin{rem}
We remark that soliton solutions of (\ref{dlylv}) are also
obtained by Hirota's direct method \cite{Hirota-direct}, which does not require any reduction.
We show a few examples of deriving soliton solutions without using the soliton solutions of
the discrete \kp.
Let $f_n(t)$ be 1-soliton or 2-soliton with the perturbation parameter $\eps$:
\begin{eqnarray}
    \label{1soliton}
\fl    f_n(t)
    =1+\eps P_1^ne^{Q_1t+\xi_{01}}\,,\\
    \label{2soliton}
\fl    f_n(t)
    =1+\eps P_1^ne^{Q_1t+\xi_{01}}+\eps P_2^ne^{Q_2t+\xi_{02}}
    +\eps^2 a_{12}(P_1P_2)^ne^{(Q_1+Q_2)t+\xi_{01}+\xi_{02}}\,.
\end{eqnarray}
Substituting them into the bilinear equations (\ref{dlylv2})
and assuming that each of the orders of $\eps$ vanishes, we have the following conditions.
\begin{equation}
    F(P_i,Q_i)=0\quad (i=1,2)\,,\qquad
    a_{12}=-\frac{F(P_1/P_2,Q_1-Q_2)}{F(P_1P_2,Q_1+Q_2)}\,,
\end{equation}
where the function $F$ is defined by
\begin{equation}
\fl    F(P,Q)
    =Q\left(\sqrt{P}e^{\tau Q}-\frac{1}{\sqrt{P}}e^{-\tau Q}\right)
    -\left(\sqrt{P}-\frac{1}{\sqrt{P}}\right)
    \left(Pe^{\tau Q}-\frac{1}{P}e^{-\tau Q}\right)\,.
\end{equation}
By computations, we can check that
the solutions (\ref{1soliton}) and (\ref{2soliton}) are equivalent
to the case of $N=1,2$ in the $N$-soliton solution (\ref{dlylv_sol}).
As we can see from this example,
we can use Hirota's direct method even if the bilinear equation includes some delays.
\end{rem}

Let us move the discussion to the nonlinear form of the delay LV equation.
Via the dependent variable transformation
\begin{equation}
    u_{n}(t)
    =\frac{f_{n+1}(t+\tau)f_{n-2}(t-\tau)}{f_{n}(t+\tau)f_{n-1}(t-\tau)}\,,
\end{equation}
the bilinear equation (\ref{dlylv}) is transformed into the nonlinear \dlydiff\ equation
\begin{equation}
    \label{dlylv_nl}
\fl     \od{}{t}\log\frac{u_{n}(t+\tau)}{u_{n-1}(t-\tau)}
    =u_{n+1}(t+\tau)-u_{n}(t+\tau)-u_{n-1}(t-\tau)+u_{n-2}(t-\tau)\,,
\end{equation}
which is the nonlinear form of (\ref{dlylv}).
If $\tau=0$, equation (\ref{dlylv_nl}) is just a subtraction of
the following nonlinear forms of the LV equation~\cite{Hirota-LV}:
\begin{equation}
    \fl\od{}{t}\log u_{n}(t)=u_{n+1}(t)-u_{n-1}(t)\,,\qquad
    \od{}{t}\log u_{n-1}(t)=u_{n}(t)-u_{n-2}(t)\,.
\end{equation}

Next, we derive the bilinear form of the delay LV equation (\ref{dlylv}) and
the $N$-soliton solution (\ref{dlylv_sol})
directly from the semi-discrete KP equation~\cite{Li-semi_discrete_KP}
\begin{equation}
\label{semidiskp}
    acD_{t}f_{n+1}^{k}(t)\cdot f_{n}^{k+1}(t)
    -(a-c)(f_{n+1}^{k+1}(t)f_{n}^{k}(t)-f_{n+1}^{k}(t)f_{n}^{k+1}(t))
    =0\,.
\end{equation}
We remark that the bilinear equation (\ref{semidiskp}) and its solutions are
derived by the continuum limit $b\to0$, $mb=t$ of the discrete \kp\ (\ref{dkp}) and
its solutions.

The Wronskian solution of the semi-discrete KP equation (\ref{semidiskp})
is given as follows~\cite{Li-semi_discrete_KP}:
\begin{eqnarray}
\label{semidiskp_sol}
&&\fl f_{n}^{k}(t)=
\left|
\begin{array}{cccc}
\phi_1(n,k,t) & \phi_1^{(1)}(n,k,t) & \cdots & \phi_1^{(N-1)}(n,k,t)\\
\phi_2(n,k,t) & \phi_2^{(1)}(n,k,t) & \cdots & \phi_2^{(N-1)}(n,k,t)\\
\vdots & \vdots & \cdots & \vdots \\
\phi_N(n,k,t) & \phi_N^{(1)}(n,k,t) & \cdots & \phi_N^{(N-1)}(n,k,t)
\end{array}
\right|\,,
\\
&&\fl \frac{\phi_i(n,k,t)-\phi_i(n-1,k,t)}{a}=\pd{\phi_i(n,k,t)}{t}\,,\quad
\frac{\phi_i(n,k,t)-\phi_i(n,k-1,t)}{c}=\pd{\phi_i(n,k,t)}{t}\,,\nonumber\\
&&\fl \phi_i^{(l)}(n,k,t)\equiv \frac{\partial^l \phi_i(n,k,t)}{\partial t^l}\,.\nonumber
\end{eqnarray}

The $N$-soliton solution of the semi-discrete KP equation is given by choosing the elements in the Wronskian solution as
\begin{eqnarray}
\label{semidkp:solitonsol}
\fl \phi_i(n,k,t)=(1-ap_i)^{-n}(1-cp_i)^{-k}e^{p_it+\zeta_{0i}}
+(1-aq_i)^{-n}(1-cq_i)^{-k}e^{q_it+\eta_{0i}}\,.\\
(i=1,2,\ldots,N)\nonumber
\end{eqnarray}

Now, we can obtain the bilinear equation (\ref{dlylv})
by applying the reduction condition
\begin{equation}
    f_{n}^{k+1}(t)\Bumpeq f_{n+2}^{k}(t+2\tau)
\end{equation}
and setting $a=2$, $c=-2$ to the semi-discrete \kp\ (\ref{semidiskp}).
Here, the relation $g_{n}^{k}(t)\Bumpeq h_{n}^{k}(t)$ is defined by
\begin{equation}
    g_{n}^{k}(t)=\left(C_0C_1^nC_2^ke^{C_3t}\right) h_{n}^{k}(t)\,,\quad C_l=\const\,.\quad(l=0,1,2,3)
\end{equation}
To realize this reduction condition ($C_1=C_2=1,C_3=0$) for  the $N$-soliton solutions, we can apply the constraint
\begin{equation}
\label{reduction2}
(1-ap_i)^{-2}(1-cp_i)e^{2\tau p_i}=(1-aq_i)^{-2}(1-cq_i)e^{2\tau q_i}
\end{equation}
to (\ref{semidkp:solitonsol}).
Replacing $p_i$ and $q_i$ by $(-2p_i-1)/a$ and $(-2q_i-1)/a$
respectively and setting $a=2$, $c=-2$, we obtain
the $N$-soliton solution of the delay LV equation in the Wronskian form
\begin{eqnarray}
\label{dlylv_sol:wr}
&&\fl f_{n}(t)=
\left|
\begin{array}{cccc}
\phi_1(n,t) & \phi_1^{(1)}(n,t) & \cdots & \phi_1^{(N-1)}(n,t)\\
\phi_2(n,t) & \phi_2^{(1)}(n,t) & \cdots & \phi_2^{(N-1)}(n,t)\\
\vdots & \vdots & \cdots & \vdots \\
\phi_N(n,t) & \phi_N^{(1)}(n,t) & \cdots & \phi_N^{(N-1)}(n,t)
\end{array}
\right|\,,\\
&&\fl  \phi_i(n,t)=(1+p_i)^{-n}e^{-p_it+\tilde{\zeta}_{0i}}
+(1+q_i)^{-n}e^{-q_it+\tilde{\eta}_{0i}}
\,,\qquad (i=1,2,\ldots,N)\nonumber\\
&&
\fl \frac{q_i}{p_i}=\left(\frac{1+q_i}{1+p_i}\right)^{2}e^{2\tau (q_i-p_i)}\nonumber\,.
\end{eqnarray}

\begin{rem}
We can also obtain the delay LV equation from
the B\"acklund transformation (BT) of the 2DTL equation~\cite{BT2DTL,Hirota-direct}
\begin{equation}
\label{BTof2DTL}
\mu D_tf_{n}^{k-1}(t)\cdot f_{n+1}^{k}(t)
-f_{n}^{k-1}(t)f_{n+1}^{k}(t)+f_{n+1}^{k-1}(t)f_n^{k}(t)=0\,.
\end{equation}
The following derivation is much easier than the above derivation from the
semi-discrete KP equation.

The $N$-soliton solution in the Wronskian (Casorati determinant)
form is given as
\begin{eqnarray}
&&\fl f_{n}^{k}(t)=
\left|
\begin{array}{cccc}
\phi_1(n,k,t) & \phi_1^{(1)}(n,k,t) & \cdots & \phi_1^{(N-1)}(n,k,t)\\
\phi_2(n,k,t) & \phi_2^{(1)}(n,k,t) & \cdots & \phi_2^{(N-1)}(n,k,t)\\
\vdots & \vdots & \cdots & \vdots \\
\phi_N(n,k,t) & \phi_N^{(1)}(n,k,t) & \cdots & \phi_N^{(N-1)}(n,k,t)
\end{array}
\right|\,,\\
&&\fl  \phi_i(n,k,t)=p_i^k(1-\mu p_i)^{-n}e^{-p_it+\zeta_{0i}}
+q_i^k(1-\mu q_i)^{-n}e^{-q_it+\eta_{0i}}
\,.\qquad (i=1,2,\ldots,N)\nonumber\label{BT2DTL:solitonsol}
\end{eqnarray}
Applying the reduction condition
\begin{equation}
    f_{n}^{k-1}(t)\Bumpeq f_{n+2}^{k}(t+2\tau)
\end{equation}
and setting $\mu=-1$, we obtain the bilinear equation (\ref{dlylv})
and its $N$-soliton solution (\ref{dlylv_sol:wr}).
\end{rem}

\end{section}

\begin{section}{An integrable delay \toda\ equation}

In this section, we construct a delay differential-difference equation
that should be called an integrable delay TL equation by using our method.

We first apply the reduction condition
\begin{equation}
    f_{n,m,k+1}=f_{n+1,m+1+\al,k}
\end{equation}
to the discrete \kp\ (\ref{dkp}) and its $N$-soliton solution (\ref{dkp_sol}).
The independent variables $n$ and $m$ are considered to be the discrete space variable
and discrete time variable respectively,
and the parameter $\alpha$ is a delay parameter.
Replacing $p_i$ and $q_i$ by $(1-p_i)/c$ and $(1-q_i)/c$ respectively and
setting $\al\del=(a-c)/a$, $\del=b/(b-c)$, we have
\begin{equation}
    \label{dlydistoda}
    f_{n}^{m+1+\al}f_{n}^{m-1}
    -\al\del^2f_{n+1}^{m+\al}f_{n-1}^{m}
    -(1-\al\del^2)f_{n}^{m+\al}f_{n}^{m}
    =0\,,
\end{equation}
\begin{eqnarray}
    \label{dlydistoda_sol}
\fl f_{n}^{m}&=&\det \left(\delta_{ij}+
\frac{\Phi_j}{p_i-q_j}\right)_{1\leq i,j\leq N}
    =\sum_{I\subset\{1,\ldots,N\}}
    \prod_{i\in I}\Phi_i
    \prod_{i<j,\ i,j\in I}\frac{(p_i-p_j)(q_i-q_j)}{(p_i-q_j)(q_i-p_j)}\,,\\
\fl\Phi_i&=&\beta_i \gamma_i
\bfrac{q_i-\al\del}{p_i-\al\del}^{n}\bfrac{1-\del q_i}{1-\del p_i}^{m}\,,\nonumber\\
\fl\frac{q_i}{p_i}
    &=&\frac{q_i-\al\del}{p_i-\al\del}\bfrac{1-\del q_i}{1-\del p_i}^{1+\al}\,.\nonumber
\end{eqnarray}
The bilinear equation (\ref{dlydistoda}) is rewritten as
\begin{eqnarray}
    \fl\left(2\sinh\bfrac{D_m}{2}\sinh\left(\frac{D_m}{2}+\frac{\al D_m}{2}\right)
    -2\al\del^2\sinh\bfrac{D_n}{2}\sinh\left(\frac{D_n}{2}+\frac{\al D_m}{2}\right)\right)\nonumber
    f_{n}^{m}\cdot f_{n}^{m}\\=0\,.
\end{eqnarray}
We can consider that equation (\ref{dlydistoda}) is the bilinear equation of the delay-difference analogue of the TL equation.
We call this the delay discrete TL equation.

By the dependent variable transformation
\begin{equation}
    1+V_{n}^{m}
    =\frac{f_{n+1}^{m+\al}f_{n-1}^{m}}{f_{n}^{m+\al}f_{n}^{m}}\,,
\end{equation}
we can transform (\ref{dlydistoda}) into the nonlinear delay-difference equation
\begin{equation}
    \label{dlydistoda_nl}
    \frac{(1+V_{n}^{m+1+\al})(1+V_{n}^{m-1})}{(1+V_{n}^{m+\al})(1+V_{n}^{m})}
    =\frac{(1+\al\del^2V_{n+1}^{m+\al})(1+\al\del^2V_{n-1}^{m})}{(1+\al\del^2V_{n}^{m+\al})(1+\al\del^2V_{n}^{m})}\,,
\end{equation}
which is the delay discrete TL equation.

Now, we apply the \dlydiff\ limit
\begin{equation}
    \del\to0\,,\qquad
    m\del=t\,,\qquad
    \al\del=2\tau
\end{equation}
to (\ref{dlydistoda}) and (\ref{dlydistoda_sol}),
where $t$ is the continuous time variable and $\tau$ is the delay parameter.
Then we obtain
\begin{eqnarray}
    \label{dlytoda}
    \fl\hd{}{t}{f_{n}(t+\tau)}{f_{n}(t-\tau)}
    -2\tau(f_{n+1}(t+\tau)f_{n-1}(t-\tau)-f_{n}(t+\tau)f_{n}(t-\tau))
    =0
\end{eqnarray}
and
\begin{eqnarray}
\label{dlytoda_sol}
\fl f_{n}(t)
&=&
\det \left(\delta_{ij}+
\frac{\Phi_j}{p_i-q_j}\right)_{1\leq i,j\leq N}
=\sum_{I\subset\{1,\ldots,N\}}
    \prod_{i\in I}\Phi_i
    \prod_{i<j,\ i,j\in I}\frac{(p_i-p_j)(q_i-q_j)}{(p_i-q_j)(q_i-p_j)}\,,\\
\fl \Phi_i&=&\beta_i \gamma_i \bfrac{q_i-2\tau}{p_i-2\tau}^{n}e^{(p_i-q_i)t}\,,\nonumber\\
\fl \frac{q_i}{p_i}
    &=&\frac{q_i-2\tau}{p_i-2\tau}e^{2\tau(p_i-q_i)}\,.\nonumber
\end{eqnarray}
Here the bilinear equation (\ref{dlytoda}) is equivalent to
\begin{equation}
\fl \left(D_t\sinh\left(\tau D_t\right)-4\tau\sinh\bfrac{D_n}{2}\sinh\left(\frac{D_n}{2}+\tau D_t\right)\right)
    f_{n}(t)\cdot f_{n}(t)=0\,.
\end{equation}
Calculating the limit of this equation as $\tau\to0$, we obtain the bilinear TL
equation~\cite{Hirota-discreteToda}
\begin{equation}
    \left(D^2_t-4\sinh^2\bfrac{D_n}{2}\right)
    f_{n}(t)\cdot f_{n}(t)=0\,.
\end{equation}
The last relation of (\ref{dlytoda_sol}) is described by $(p_i+1/p_i-q_i-1/q_i)\tau+O(\tau^2)=0$,
thus we have $p_i=1/q_i$ as $\tau\to0$.
Therefore, we can check that the limit of (\ref{dlytoda_sol}) as $\tau\to0$ is
the $N$-soliton solution of the TL equation~\cite{Hirota-direct,Hirota-Satsuma-Toda,Hirota-discreteToda}.
Thus we claim that equation
(\ref{dlytoda}) is the bilinear equation of the delay TL equation.

\begin{rem}
If we use the relation $\del=(a-c)/a$ instead of
the above one $\al\del=(a-c)/a$, we obtain the bilinear equation
\begin{equation*}
    f_{n}^{m+1+\al}f_{n}^{m-1}
    -\del^2f_{n+1}^{m+\al}f_{n-1}^{m}
    -(1-\del^2)f_{n}^{m+\al}f_{n}^{m}
    =0
\end{equation*}
instead of the delay discrete TL equation (\ref{dlydistoda}).
This equation can be considered more natural than (\ref{dlydistoda}),
because we can obtain the discrete TL equation from it by
putting $\al=0$~\cite{Hirota-discreteToda}.
However it does not yield a good \dlydiff\ equation,
because the order $O(\del^2)$ vanishes in the \dlydiff\ limit.
On the other hand, the bilinear equation (\ref{dlydistoda}) yields
the good \dlydiff\ equation (\ref{dlytoda}), which should be called a delay TL equation.
\end{rem}

We present the nonlinear form of the bilinear equation of
the delay TL equation (\ref{dlytoda})
under the dependent variable transformation
\begin{equation}
    1+V_{n}(t)
    =\frac{f_{n+1}(t+\tau)f_{n-1}(t-\tau)}{f_{n}(t+\tau)f_{n}(t-\tau)}\,.
\end{equation}
By using this transformation,
we can transform (\ref{dlytoda}) into the \dlydiff\ equation
\begin{eqnarray}
    \label{dlytoda_nl}
    \fl\bfrac{1}{2\tau}\od{}{t}\log\frac{1+V_{n}(t+\tau)}{1+V_{n}(t-\tau)}\nonumber\\
    =V_{n+1}(t+\tau)+V_{n-1}(t-\tau)-V_{n}(t+\tau)-V_{n}(t-\tau)\,,
\end{eqnarray}
which is the nonlinear form of (\ref{dlytoda}).
The limit of (\ref{dlytoda_nl}) as $\tau\to0$ is the nonlinear form of the TL
equation~\cite{Hirota-direct,Hirota-Satsuma-Toda,Hirota-discreteToda}:
\begin{equation}
    \od[2]{}{t}\log(1+V_{n}(t))=V_{n+1}(t)-2V_{n}(t)+V_{n-1}(t)\,.
\end{equation}

We can also obtain the above result from the BT of the 2DTL equation (\ref{BTof2DTL}):
\begin{equation}
\mu D_tf_{n}^{k-1}(t)\cdot f_{n+1}^{k}(t)
-f_{n}^{k-1}(t)f_{n+1}^{k}(t)+f_{n+1}^{k-1}(t)f_n^{k}(t)=0\,.
\end{equation}
Applying the reduction condition
\begin{equation}
    f_{n}^{k-1}(t)\Bumpeq f_{n+1}^{k}(t+2\tau)
\end{equation}
and setting $\mu=-1/(2\tau)$, we obtain the bilinear equation (\ref{dlytoda})
and its $N$-soliton solution
\begin{eqnarray}
&&\fl f_{n}(t)=
\left|
\begin{array}{cccc}
\phi_1(n,t) & \phi_1^{(1)}(n,t) & \cdots & \phi_1^{(N-1)}(n,t)\\
\phi_2(n,t) & \phi_2^{(1)}(n,t) & \cdots & \phi_2^{(N-1)}(n,t)\\
\vdots & \vdots & \cdots & \vdots \\
\phi_N(n,t) & \phi_N^{(1)}(n,t) & \cdots & \phi_N^{(N-1)}(n,t)
\end{array}
\right|\,,\\
&&\fl  \phi_i(n,t)=(2\tau+ p_i)^{-n}e^{-p_it+\tilde{\zeta}_{0i}}
+(2\tau+ q_i)^{-n}e^{-q_it+\tilde{\eta}_{0i}}
\,,\qquad (i=1,2,\ldots,N)\nonumber\\
&&\fl \frac{q_i}{p_i}=\frac{2\tau +q_i}{2\tau +p_i}e^{2\tau(q_i-p_i)}\,,\nonumber
\end{eqnarray}
which leads to (\ref{dlytoda_sol}) by replacing $p_i$ and $q_i$ by $-p_i$ and $-q_i$.
This construction of the delay TL equation
does not require the delay-differential limit.

\end{section}

\begin{section}{An integrable delay \sg\ equation}

In this section, we find an integrable delay sG equation by the process similarly to the previous sections. It is a delay partial differential equation which can be obtained simply.

We consider the bilinear equation of the discrete 2DTL
equation~\cite{Hirota-discrete_KP,Hirota-discrete2DTL}
\begin{eqnarray}
    \fl abf_{k+1,n+1,m}f_{k-1,n,m+1}
    +f_{k,n+1,m}f_{k,n,m+1}
    -(1+ab)f_{k,n+1,m+1}f_{k,n,m}=0\,,\label{d2DTL}
\end{eqnarray}
where $k$ is the discrete space variable, and $n,m$ are the discrete time variables.
The $N$-soliton solution of (\ref{d2DTL}) is given as follows~\cite{Hirota-discrete2DTL}:
\begin{eqnarray}
    \label{d2DTL_sol}
\fl &&f_{k,n,m}=
\det \left(\delta_{ij}+\frac{\phi_i\psi_j}{p_i-q_j}\right)_{1\leq i,j\leq N}\\
 \fl && \hspace{1cm} =\det \left(\delta_{ij}+\frac{\phi_j \psi_j}
{p_i-q_j}\right)_{1\leq i,j\leq N}
=\sum_{I\subset\{1,\ldots,N\}}\prod_{i\in I}\Phi_i
    \prod_{i<j,\ i,j\in I}\frac{(p_i-p_j)(q_i-q_j)}{(p_i-q_j)(q_i-p_j)}\,,\nonumber\\
\fl &&\phi_i=\beta_ip_i^{k}(1-a p_i)^{-n}(1+b/p_i)^{-m}
\,,\quad \psi_i=\gamma_iq_i^{-k}(1-a q_i)^{n}(1+b/q_i)^{m} \,,\nonumber\\
\fl &&\Phi_i
    =\beta_i\gamma_i\bfrac{p_i}{q_i}^{k}
\bfrac{1-aq_i}{1-ap_i}^{n}\bfrac{1+b/q_i}{1+b/p_i}^{m}\,.\nonumber
\end{eqnarray}
Applying the reduction condition
\begin{equation}
 f_{k+1,n-\alpha,m-\beta}=f_{k-1,n+\alpha,m+\beta}\,,
\end{equation}
and setting
\begin{equation}
\fl f_n^m\equiv f_{k,n,m}=f_{k-2,n+2\alpha,m+2\beta}\,,\quad
g_{n}^{m}\equiv f_{k+1,n-\alpha,m-\beta}=f_{k-1,n+\alpha,m+\beta}\,,\quad a=b=\delta
\end{equation}
to (\ref{d2DTL}) and (\ref{d2DTL_sol}) respectively, we obtain
\begin{eqnarray}
 &&(1+\del^2)f_{n+1}^{m+1}f_{n}^{m}
    -f_{n+1}^{m}f_{n}^{m+1}
    -\del^2g_{n+1+\al}^{m+\be}g_{n-\al}^{m+1-\be}
    =0\,, \label{dlydissg1-1}\\
&&(1+\del^2)g_{n+1}^{m+1}g_{n}^{m}
    -g_{n+1}^{m}g_{n}^{m+1}
    -\del^2f_{n+1+\al}^{m+\be}f_{n-\al}^{m+1-\be}
    =0 \label{dlydissg1-2}
\end{eqnarray}
and
\begin{eqnarray}
    \label{dlydissg_sol}
    &&f_{n}^{m}=\det \left(\delta_{ij}+\frac{\Phi_j}
{p_i-q_j}\right)_{1\leq i,j\leq N}\,,\\
    &&\Phi_i
    =\beta_i\gamma_i\bfrac{1-\del q_i}{1-\del p_i}^{n}\bfrac{1+\del/q_i}{1+\del/p_i}^{m}\,,\nonumber\\
    &&\bfrac{p_i}{q_i}^2
    =\bfrac{1-\del q_i}{1-\del p_i}^{2\al}
\bfrac{1+\del/q_i}{1+\del/p_i}^{2\be}\,.\nonumber
\end{eqnarray}
To construct a delay-difference analogue of the sG equation, we take $g_n^m$
to be the complex conjugate of $f_n^m$.
Considering the regularity conditions $f_n^m\neq 0$,
the $N$-soliton solution is given as
\begin{eqnarray}
    \label{dlydissg_sol2}
&&\fl f_{n}^{m}=\det \left(\delta_{ij}+\frac{\Phi_j}
{p_i-q_j}\right)_{1\leq i,j\leq N}\,, \quad
g_{n}^{m}=\det \left(\delta_{ij}-\frac{\Phi_j}{p_i-q_j}\right)_{1\leq i,j\leq N}\,,\\
&&\fl \Phi_i
    =\sqrt{-1}\mu_i
\bfrac{1-\del q_i}{1-\del p_i}^{n}\bfrac{1+\del/q_i}{1+\del/p_i}^{m}\,,
\nonumber\\
    &&\fl \frac{p_i}{q_i}
    =-\bfrac{1-\del q_i}{1-\del p_i}^{\al}
\bfrac{1+\del/q_i}{1+\del/p_i}^{\be}\,,\nonumber
\end{eqnarray}
where $\mu_i$ is a real constant.

Setting $f_{n}^{m}=F_{n}^{m}+\sqrt{-1}G_{n}^{m}$ and
$g_{n}^{m}=F_{n}^{m}-\sqrt{-1}G_{n}^{m}$, we can rewrite
the bilinear equations (\ref{dlydissg1-1}) and (\ref{dlydissg1-2}) with
\begin{eqnarray}
    &&\fl\left(2\sinh\bfrac{D_n}{2}\sinh\bfrac{D_m}{2}
    +2\del^2\sinh\left(\frac{D_n}{2}+\frac{\al D_n}{2}+\frac{\be D_m}{2}\right)
    \sinh\left(\frac{D_m}{2}-\frac{\al D_n}{2}-\frac{\be D_m}{2}\right)\right)\nonumber\\
    &&(F_{n}^{m}\cdot F_{n}^{m}-G_{n}^{m}\cdot G_{n}^{m})=0\,, \label{dlydissg2-1}\\
    &&\fl\left(2\sinh\bfrac{D_n}{2}\sinh\bfrac{D_m}{2}
    +2\del^2\cosh\left(\frac{D_n}{2}+\frac{\al D_n}{2}+\frac{\be D_m}{2}\right)
    \cosh\left(\frac{D_m}{2}-\frac{\al D_n}{2}-\frac{\be D_m}{2}\right)\right)\nonumber\\
    &&F_{n}^{m}\cdot G_{n}^{m}=0\,.  \label{dlydissg2-2}
\end{eqnarray}
When $\al=\be=0$, we can find that
the bilinear equations (\ref{dlydissg1-1}), (\ref{dlydissg1-2}) (and also
(\ref{dlydissg2-1}), (\ref{dlydissg2-2}))
and the $N$-soliton solution (\ref{dlydissg_sol2}) are actually
equivalent to the bilinear equations of the discrete sG equation and their
$N$-soliton solution~\cite{Hirota-discrete_sG}.

To construct a nonlinear form of the delay-difference analogue of the sG equation, we
consider the dependent variable transformation
\begin{equation}
f_{n}^{m}=\exp\left(\frac{\rho_{n}^{m}}{4}+\sqrt{-1}\frac{\theta_{n}^{m}}{4}\right)\,,
\quad
g_{n}^{m}=\exp\left(\frac{\rho_{n}^{m}}{4}-\sqrt{-1}\frac{\theta_{n}^{m}}{4}\right)\,,
\end{equation}
which is equivalent to
\begin{equation}
\fl\theta_n^m=2\sqrt{-1}\log \frac{g_n^m}{f_n^m}=4 \tan^{-1}\frac{G_n^m}{F_n^m}\,,
\quad
\rho_n^m=2\log f_n^mg_n^m=2\log((F_n^m)^2+(G_n^m)^2)\,.
\end{equation}
By using this transformation,
we can transform the bilinear equations (\ref{dlydissg1-1}) and (\ref{dlydissg1-2})
into the nonlinear
delay-difference equation
\begin{eqnarray}
 &&\fl\sin\left(\frac{\theta_{n+1}^{m}+\theta_{n}^{m+1}-\theta_{n+1}^{m+1}-\theta_{n}^{m}}{4}\right)\nonumber\\
&&\fl=\del^2\exp\left(\frac{\rho_{n+1+\al}^{m+\be}+\rho_{n-\al}^{m+1-\be}-\rho_{n+1}^{m}-\rho_{n}^{m+1}}{4}\right)
    \sin\left(\frac{\theta_{n+1+\al}^{m+\be}+\theta_{n-\al}^{m+1-\be}+\theta_{n+1}^{m+1}+\theta_{n}^{m}}{4}\right)\,,\nonumber \\
&&\label{dlydissg_nl-1}\\
&&\fl \sinh\left(\frac{\rho_{n+1}^{m}+\rho_{n}^{m+1}-\rho_{n+1}^{m+1}-\rho_{n}^{m}}{4}\right)
=\del^2\exp\left(\frac{-\rho_{n+1}^{m}-\rho_{n}^{m+1}+\rho_{n+1}^{m+1}+\rho_{n}^{m}}{4}\right)
 \nonumber
\\
 &&\fl-\del^2\exp\left(\frac{\rho_{n+1+\al}^{m+\be}+\rho_{n-\al}^{m+1-\be}-\rho_{n+1}^{m+1}-\rho_{n}^{m}}{4}\right)
    \cos\left(\frac{\theta_{n+1+\al}^{m+\be}+\theta_{n-\al}^{m+1-\be}+\theta_{n+1}^{m}+\theta_{n}^{m+1}}{4}\right)\nonumber\\
&&\fl -\del^4\exp\left(\frac{\rho_{n+1+\al}^{m+\be}+\rho_{n-\al}^{m+1-\be}-\rho_{n+1}^{m}-\rho_{n}^{m+1}}{4}\right)
    \sinh\left(\frac{\rho_{n+1+\al}^{m+\be}+\rho_{n-\al}^{m+1-\be}-\rho_{n+1}^{m+1}-\rho_{n}^{m}}{4}\right)\,,  \nonumber\\
 \label{dlydissg_nl-2}
\end{eqnarray}
which is the delay-difference analogue of the sG equation.
Equation (\ref{dlydissg_nl-1}) in the case of $\al=\be=0$ is
the discrete sG equation~\cite{Hirota-discrete_sG}
\begin{equation}
    \fl\sin\left(\frac{\theta_{n+1}^{m}+\theta_{n}^{m+1}-\theta_{n+1}^{m+1}-\theta_{n}^{m}}{4}\right)
    =\del^2\sin\left(\frac{\theta_{n+1}^{m}+\theta_{n}^{m+1}+\theta_{n+1}^{m+1}+\theta_{n}^{m}}{4}\right)\,.
\end{equation}

Now, let us apply the \dlydiff\ limit
\begin{equation}
    \del\to0\,,\quad
    n\del=x\,,\quad
    \al\del=\xi\,,\quad
    m\del=y\,,\quad
    \be\del=\eta
\end{equation}
to the bilinear equations (\ref{dlydissg1-1}), (\ref{dlydissg1-2})
and the $N$-soliton solution (\ref{dlydissg_sol2}).
Here $x,y$ are the continuous variables, and $\xi,\eta$ are the delay parameters.
Consequently we obtain the bilinear equations
\begin{eqnarray}
&& \fl D_x\hd{}{y}{f(x,y)}{f(x,y)}
    +2(f(x,y)f(x,y)
    -g(x+\xi,y+\eta)g(x-\xi,y-\eta))
    =0\,,\label{dlysg1-1}\\
&& \fl D_x\hd{}{y}{g(x,y)}{g(x,y)}
    +2(g(x,y)g(x,y)
    -f(x+\xi,y+\eta)f(x-\xi,y-\eta))
    =0\,,\label{dlysg1-2}
\end{eqnarray}
and the $N$-soliton solution
\begin{eqnarray}
    \label{dlysg_sol}
&&\fl f(x,y)=\det \left(\delta_{ij}+\frac{\Phi_j}
{p_i-q_j}\right)_{1\leq i,j\leq N}\,, \quad
g(x,y)=\det \left(\delta_{ij}-\frac{\Phi_j}{p_i-q_j}\right)_{1\leq i,j\leq N}\,,\\
&&\fl \Phi_i
    =\sqrt{-1}\mu_i\exp\left((p_i-q_i)x-\left(\frac{1}{p_i}-\frac{1}{q_i}\right)y\right)\,,
\nonumber\\
 &&\fl   \frac{p_i}{q_i}
    =-\exp\left(\xi(p_i-q_i)-\eta\left(\frac{1}{p_i}-\frac{1}{q_i}\right)\right)\,.
\nonumber
\end{eqnarray}

Setting $f(x,y)=F(x,y)+\sqrt{-1}G(x,y)$ and $g(x,y)=F(x,y)-\sqrt{-1}G(x,y)$,
we can rewrite (\ref{dlysg1-1}) and (\ref{dlysg1-2})
with
\begin{eqnarray}
&& \fl\left(D_xD_y
    -4\sinh^2\left(\frac{\xi D_x}{2}+\frac{\eta D_y}{2}\right)
    \right)(F\cdot F-G\cdot G)=0\,,    \label{dlysg2-1}\\
&& \fl\left(D_xD_y
    +4\cosh^2\left(\frac{\xi D_x}{2}+\frac{\eta D_y}{2}\right)
    \right)F\cdot G=0\,.    \label{dlysg2-2}
\end{eqnarray}
In the case of $\xi=\eta=0$, the bilinear equations
(\ref{dlysg1-1}), (\ref{dlysg1-2}) (and also (\ref{dlysg2-1}), (\ref{dlysg2-2}))
and the $N$-soliton solution (\ref{dlysg_sol}) lead to
the sG equation and their $N$-soliton
solution~\cite{Hirota-discrete_sG,Hirota-sG,Ablowitz-Segur}.

To construct a nonlinear form of the delay-differential analogue of the sG equation, we
consider the dependent variable transformation
\begin{equation}
\fl f(x,y)=\exp\left(\frac{\rho(x,y)}{4}+\sqrt{-1}\frac{\theta(x,y)}{4}\right)\,, \quad
g(x,y)=\exp\left(\frac{\rho(x,y)}{4}-\sqrt{-1}\frac{\theta(x,y)}{4}\right)\,,
\end{equation}
which is equivalent to
\begin{eqnarray}
\fl\theta(x,y)=2\sqrt{-1}\log \frac{g}{f}=4\tan^{-1}\frac{G}{F}\,,
\quad
\rho(x,y)=2\log fg=2\log(F^2+G^2)\,.
\end{eqnarray}
By using this transformation,
we can transform the bilinear equations
(\ref{dlysg1-1}) and (\ref{dlysg1-2}) into the nonlinear \dlydiff\ equation
\begin{eqnarray}
&& \fl\frac{\partial^2}{\partial x\partial y}\theta(x,y)
    =-4\exp\left(\frac{\rho(x+\xi,y+\eta)+\rho(x-\xi,y-\eta)-2\rho(x,y)}{4}\right)\nonumber\\
&&\times\sin\left(\frac{\theta(x+\xi,y+\eta)+\theta(x-\xi,y-\eta)+2\theta(x,y)}{4}\right)\,,
\label{dlysg_nl-1}\\
&& \fl\frac{\partial^2}{\partial x\partial y}\rho(x,y)
=-4+4\exp\left(\frac{\rho(x+\xi,y+\eta)+\rho(x-\xi,y-\eta)-2\rho(x,y)}{4}\right)\nonumber\\
&&\times\cos\left(\frac{\theta(x+\xi,y+\eta)+\theta(x-\xi,y-\eta)+2\theta(x,y)}{4}\right)\,,
\label{dlysg_nl-2}
\end{eqnarray}
which is the delay-differential analogue of the sG equation.
In the case of $\xi=\eta=0$, the above nonlinear equation
leads to the sG equation~\cite{Hirota-discrete_sG,Hirota-sG}:
\begin{equation}
    \frac{\partial^2}{\partial x\partial y}\theta(x,y)=-4\sin\theta(x,y)\,.
\end{equation}

We can obtain the above delay-differential analogue of the sG equation
by a reduction of the 2DTL equation .
The bilinear equation of the 2DTL equation
\begin{equation}
\fl  D_xD_y f_{k}(x,y)\cdot f_{k}(x,y)
    +2(f_{k}(x,y)f_{k}(x,y)-f_{k+1}(x,y)f_{k-1}(x,y))
    =0
\end{equation}
has the following $N$-soliton solution~\cite{Hirota-direct,Hirota-discrete2DTL}:
\begin{eqnarray}
\fl &&f_k(x,y)=\det \left(\delta_{ij}+\frac{\phi_i \psi_j}{p_i-q_j}\right)_{1\leq i,j\leq N}\\
\fl && \hspace{1.3cm} =\det \left(\delta_{ij}+\frac{\phi_j \psi_j}
{p_i-q_j}\right)_{1\leq i,j\leq N}
=\sum_{I\subset\{1,\ldots,N\}}\prod_{i\in I}\Phi_i
\prod_{i<j,\ i,j\in I}\frac{(p_i-p_j)(q_i-q_j)}{(p_i-q_j)(q_i-p_j)}\,,\nonumber\\
\fl &&\phi_i=
  \beta_i p_i^ke^{p_ix-p_i^{-1}y}\,, \quad
   \psi_i =\gamma_iq_i^{-k}
    e^{-q_ix+q_i^{-1}y}\,,\nonumber\\
\fl && \Phi_i
    =\beta_i\gamma_i
\bfrac{p_i}{q_i}^{k}\exp\left((p_i-q_i)x-\left(\frac{1}{p_i}-\frac{1}{q_i}\right)y\right)\,,\nonumber
\end{eqnarray}
where $p_i, q_i, \beta_i, \gamma_i$ are constants.
We apply the reduction condition
\begin{equation}
f_{k+1}(x-\xi,y-\eta)=f_{k-1}(x+\xi,y+\eta) \label{SG-reduction}
\end{equation}
and set
\begin{eqnarray}
&&f(x,y)\equiv f_k(x,y)=f_{k-2}(x+2\xi,y+2\eta)\,, \\
&&g(x,y)\equiv f_{k+1}(x-\xi,y-\eta)=f_{k-1}(x+\xi,y+\eta)\,.
\end{eqnarray}
Then we obtain the bilinear equations
\begin{eqnarray}
\fl && D_xD_y f(x,y)\cdot f(x,y)
    +2(f(x,y)f(x,y)-g(x+\xi,y+\eta)g(x-\xi,y-\eta))
    =0\,,\\
\fl && D_xD_y g(x,y)\cdot g(x,y)
    +2(g(x,y)g(x,y)-f(x+\xi,y+\eta)f(x-\xi,y-\eta))
    =0\,.
\end{eqnarray}
For the $N$-soliton solution, the constraint
\begin{equation}
\left(\frac{p_i}{q_i}\right)^2
=\exp\left(2\xi (p_i-q_i)-2\eta\left(\frac{1}{p_i}-\frac{1}{q_i}\right)\right)\label{SG-reduction-solution}
\end{equation}
provides the reduction condition (\ref{SG-reduction}).
The constraint (\ref{SG-reduction-solution}) leads to
\begin{equation}
\frac{p_i}{q_i}
    =\pm\exp\left(\xi (p_i-q_i)-\eta\left(\frac{1}{p_i}-\frac{1}{q_i}\right)\right)\,.
\end{equation}
To take $g(x,y)$ to be the complex conjugate of $f(x,y)$, we can choose
\begin{equation}
\frac{p_i}{q_i}
    =-\exp\left(\xi (p_i-q_i)-\eta\left(\frac{1}{p_i}-\frac{1}{q_i}\right)\right)
\end{equation}
and
$\beta_i\gamma_i=\sqrt{-1}\mu_i$, where $\mu_i$ is a real constant.
Thus we obtain the delay-differential analogue of the sG equation and
its $N$-soliton solution from the 2DTL equation.
This construction of the delay sG equation
does not require the delay-differential limit.

\end{section}

\begin{section}{Conclusions}

We have presented the systematic method to construct delay-difference and delay-differential analogues of soliton equations and their $N$-soliton solutions.

Our construction starts from the discrete KP equation (or discrete 2DTL equation) and uses reduction and delay-differential limit.
As examples, we have obtained the delay-difference and delay-differential analogues of the LV, TL, and sG equations and their $N$-soliton solutions.
We have also presented another construction of delay-differential analogues of soliton equations starting from the semi-discrete KP, BT of 2DTL, and 2DTL equations without applying a delay-differential limit.
In the construction of them, the important thing is to integrate discrete variables with continuous variables by reduction.

In this paper, we have not discussed Lax pairs and conserved quantities of the delay soliton equations and the relationship to the delay-differential Painlev\'e equations.
These problems remain to be revealed in future studies.

This work was partially supported by JSPS KAKENHI Grant Numbers 18K03435, 17H02856, 22K03441 and JST/CREST.

\end{section}

\end{document}